\documentclass[aps,prl,twocolumn]{revtex4-1} 

\usepackage{amsmath,amssymb} 
\usepackage{bm}
\usepackage{graphicx} 
\usepackage{siunitx}

\providecommand{\vect}[1]{\boldsymbol{#1}}
\DeclareMathOperator{\sech}{sech}

\begin{document} 
\title{New Boundary-Driven Twist States in Systems with Broken Spatial Inversion Symmetry} 

\author{Kjetil M.\ D.\ Hals} 
\affiliation{Institute of Physics, Johannes Gutenberg University, 55128 Mainz, Germany\\
 Faculty of Engineering and Science, Western Norway University of Applied Sciences, No-6803 F\o rde, Norway} 

\author{Karin Everschor-Sitte} 
\affiliation{Institute of Physics, Johannes Gutenberg University, 55128 Mainz, Germany} 

\begin{abstract}
A full description of a magnetic sample includes a correct treatment of the boundary conditions (BCs).
This is in particular important in thin film systems, where even bulk properties might be modified by the properties of the boundary of the sample.
We study generic ferromagnets with broken spatial inversion symmetry and derive the general micromagnetic BCs of a system with Dzyaloshinskii-Moriya interaction (DMI). 
We demonstrate that the BCs require the full tensorial structure of the third-rank DMI tensor and not just the antisymmetric part, which is usually taken into account.
Specifically, we study systems with $C_{\infty v}$ symmetry and explore the consequences of the DMI.
Interestingly, we find that the DMI already in the simplest case of a ferromagnetic thin-film leads to a purely boundary-driven magnetic twist state at the edges of the sample.
The twist state represents a new type of DMI-induced spin structure, which is completely independent of the internal DMI field.
We estimate the size of the texture-induced magnetoresistance effect being in the range of that of domain walls.
\end{abstract}

\maketitle

Over the past few years, there has been an increasing interest in magnets where interface-induced phenomena play a major role~\cite{Hellman2016, Brataas2014, Gambardella2011}. This includes the topics of magnetic heterostructures as well as thin films, where the main effects arise from the sample's boundary. Therefore, a rigorous understanding of  the physical boundary conditions (BCs) is needed.

The strong spin-orbit coupling (SOC) and broken spatial inversion symmetry of these nanostructures lead to an intricate interplay between spin, charge, and orbital degrees of freedom, which affect the magnetic equilibrium state as well as the current-driven spin phenomena. Important examples of the SOC effects include current-driven spin-orbit torques~\cite{Bernevig2005c, Manchon2008, Chernyshov2009, Garate2009, Hals2010, Miron2010, Miron2011b, Miron2011, Fang2011, Liu2012c, Liu2012, Emori2013, Ryu2013, Garello2013, Haazen2013, Fan2013, Kurebayashi2014, Emori2014},  charge-pumping via magnetization precession~\cite{Hals2010, Tatara2013, Ciccarelli2014, Hals2015, Freimuth2015,Verba2015}, and the formation of topologically nontrivial skyrmion textures~\cite{Muhlbauer2009a, Rossler2006, Yu2010, Jonietz2010, Adams2011, Heinze2011, Yu2011b, Tonomura2012, Huang2012, Kanazawa2012, Romming2013, Muller2016a} and chiral domain walls~\cite{Thiaville2012, Emori2013, Ryu2013, Rohart2013, Pyatakov2015}, as well as the multiferroic behaviour of chiral magnets~\cite{Sergienko2006,Seki2012} and the ferroelectricity of magnetic textures~\cite{Pyatakov2015, Rojac2016}.

The underlying mechanism being responsible for chiral skyrmions and domain walls is the Dzyaloshinskii-Moriya interaction (DMI)~\cite{Dzyaloshinsky1958, Moriya1960a}. The DMI is a relativistic magnetic exchange interaction that originates from broken spatial inversion symmetry. Phenomenologically, the DMI is modeled by a free-energy density term, which is linear in the spatial variations of the magnetization. In its most general form, 
the term can be written as 
\begin{equation}
\mathcal{F}_D= D_{ijk}m_i\partial_j m_k.
\label{Eq:DMIintro}
\end{equation}
as discussed for example explicitly in Landau \& Lifshitz, Ref.~\cite{Landau1984}.

Here, $\vect{m}$ is a unit vector pointing along the magnetization $\vect{M}= M_s\vect{m}$, and $D_{ijk}$ is the DMI tensor, which is linear in the relativistic interactions. 
The particular form of the DMI tensor is determined by the point group of the system.  
Here, and in what follows,  we use the convention of a summation of repeated indices.
To avoid confusion with the frequently used terminology of denoting Bloch (N\'eel) DMI as bulk (surface)-induced DMI,
we will denote the bulk part of the sample as ``internal'' and the surface as ``boundary''. 

In the present work, we investigate how the DMI affects the magnetic equilibrium state at the boundary of the sample.
Contrary to the internal DMI field, which only depends on the antisymmetric part $\sim D_{ijk} - D_{kji}$ of the DMI tensor and which was discussed already in previous works \cite{Rohart2013, Meynell2014, Rybakov2013, Wilson2013, Leonov2015, Muller2016a, Leonov2016c}, we find that a correct treatment of the micromagnetic BCs requires the full tensorial structure of the DMI. 
Our general boundary conditions comprise novel spin phenomena, such as purely boundary-driven twist states along the high symmetry axis, which we discuss below.
Because boundary effects can dominate the physics of thin samples~\cite{Hellman2016}, we expect our results to be important for thin films in which the new BCs  might even excel the effects of the internal DMI field. 

As an important example we show the result for $C_{\infty v}$ systems due to the following reasons i) they describe well the physics of ferromagnetic heterostructures such as polycrystalline $Pt/Co/AlO_x$ systems\cite{Emori2013} and ii) to show that even in high symmetry class systems, where many DMI tensor elements vanish, the symmetric DMI does lead to an effect.
Importantly we find that three independent tensor elements determine the BCs in $C_{\infty v}$ systems, whereas only a single parameter is required to model the internal DMI field.

The consequences of the novel BCs become apparent already in a very simple example of a thin-film ferromagnet with $C_{\infty v}$ symmetry subject to an out-of-plane magnetic field. Here, the DMI-induced BCs lead to a non-negligible twist state at the edges. We give an estimate for the twist-state induced magnetoresistance showing that it could be observed via standard magnetoresistance measurements.
The boundary-driven spin phenomena become of particular importance in magnetic nanostructures, in which the boundary effects strongly influence the internal magnetic structure of the system. 
Implementing the full DMI-induced BCs will therefore be of crucial importance for a correct micromagnetic modeling of magnetic nanostructures,  which represents an essential tool for exploring future spintronic devices.  

We consider a ferromagnet with broken spatial inversion symmetry and SOC, which covers a finite region. The magnetic system is assumed to be far below the Curie temperature so that longitudinal variations of the magnetization can be disregarded. In this case, the local 
magnetization $\vect{M} (\vect{r})= M_s\vect{m} (\vect{r})$ is fully determined by the unit vector $\vect{m} (\vect{r})$, which represents the local direction of the magnetization. 
Phenomenologically, the magnetic system is determined by the free energy functional, which up to second order in the magnetization gradients is given by~\cite{Landau1984}
\begin{equation}
F[ \vect{m} ]  = \int {\rm d\vect{r}}\left[  \mathcal{F}_{\rm e} + \mathcal{F}_{\rm D} + \mathcal{F}_{\rm h} + \mathcal{F}_{\rm a} \right]. \label{Eq:F}
\end{equation}
Here, $\mathcal{F}_{\rm e}$ is the symmetric magnetic exchange interaction given by $\mathcal{F}_{\rm e}=J_{ij} \partial_i \vect{m}\cdot \partial_j \vect{m}$, where $J_{ij}$ is a symmetric positive definite matrix parameterizing the spin stiffness. $\mathcal{F}_{\rm h}= -M_s \vect{m}\cdot \vect{H}_{h}$ describes the coupling to an external magnetic field $\vect{H}_{h}$, and $\mathcal{F}_{\rm a}$ represents the anisotropy energy including dipolar interactions.
The tensorial forms of the exchange and DMI are determined by the symmetry relations~\cite{Birss1964}
\begin{align}
 J_{ij}  &= \mathcal{R}^{(\alpha)}_{il} \mathcal{R}^{(\alpha)}_{jm} J_{lm} , \label{Eq:SymRel1}\\
D_{ijk}  &=  \mathcal{R}^{(\alpha)}_{il} \mathcal{R}^{(\alpha)}_{jm} \mathcal{R}^{(\alpha)}_{kn} D_{lmn} , \label{Eq:SymRel2}
\end{align}
where $\{ \boldsymbol{\mathcal{R}}^{(\alpha)} | ~\alpha = 1, 2, ...  \}$  are the generators of the system's point group. 
Note that the tensor coefficients $D_{ijk}$ vanish for symmetry groups containing the inversion operator $\mathcal{R}_{ij} = -\delta_{ij}$, as DMI only exists in systems with spatially asymmetric SOC. 
In the literature, also different notations are used for the DMI, which we summarize and relate to our notation in the Supplemental Material.

The equilibrium state of the magnetic system is found by a variational minimization of the free energy functional  \eqref{Eq:F} with respect to small variations $\delta\vect{m} (\vect{r})$ of the local magnetization direction. 
Due to the normalization $\vect{m}\cdot\vect{m}= 1$, the variation is constrained by the condition $\delta\vect{m} (\vect{r})\cdot\vect{m} (\vect{r}) = 0$. Consequently, the local perturbation 
can be written as $\delta\vect{m} (\vect{r}) = \vect{m} (\vect{r}) \times \delta \boldsymbol{\varphi} (\vect{r})$, where $\delta \boldsymbol{\varphi}\in \mathbb{R}^3$ with $|\delta\boldsymbol{\varphi}| \ll 1$.
The equilibrium condition is determined by $\delta F[ \vect{m} (\vect{r}) ] / \delta \boldsymbol{\varphi} (\vect{r})  = 0 $ resulting in
\begin{align}
0 &= \vect{m}\times \left[ 2 J_{ij}\partial_{i}\partial_{j}\vect{m}  + \vect{H}_D + \vect{H}_a  + \vect{H}_{h} \right], \label{Eq:Eqv1}  \\
0  &=  \vect{m}\times \left[ 2 J_{ij}n_{i}\partial_{j}\vect{m}  + \boldsymbol{\Gamma}_D  \right], \label{Eq:Eqv2}
\end{align}
one equation for the inner part (Eq.~\eqref {Eq:Eqv1}) and one for the boundary of the sample (Eq.~\eqref {Eq:Eqv2}). 
The latter equation originates from the partitial derivatives in the free energy functional, which lead to surface integrals when the functional is varied with respect to $\vect{m}$.  
Consequently, the magnetic anisotropy and external magnetic field do not contribute to Eq.~\eqref {Eq:Eqv2}, but only enter the equation for the inner part of the sample via the effective fields $ \vect{H}_{h}$ and $(H_a)_k= -\partial\mathcal{F}_{\rm a}/\partial m_k$, respectively.
In contrast, the exchange interaction and the DMI enter both equations. In particular, the DMI enters via the internal DMI field $\vect{H}_D$ and the boundary-induced DMI field $\boldsymbol{\Gamma}_D$
\begin{align}
(H_{D})_k &= (D_{ijk} - D_{kji}) \partial_jm_i, \label{Eq:Hd} \\
 (\Gamma_D)_k &= m_i n_j D_{ijk}. \label{Eq:Gd}
\end{align}
Here, $\vect{n}$ is the outer surface normal of the boundary.
Eq.~\eqref{Eq:Eqv2} is satisfied when $2 J_{ij}n_{i}\partial_{j}\vect{m} $ cancels the components of $\boldsymbol{\Gamma}_D$ that are perpendicular to $\vect{m}$.
This leads to our main result of this paper, namely, the general condition for the magnetization at the boundary of the sample
\begin{equation}
2 J_{ij}n_{i}\partial_{j}\vect{m}  = - \vect{m}\times ( \boldsymbol{\Gamma}_D\times\vect{m} ) . \label{Eq:BondCond}
\end{equation} 
Note that for isotropic spin stiffness $J_{ij}= J\delta_{ij}$ and without DMI, Eq.~\eqref{Eq:BondCond} reduces to the standard Neumann BC $n_i \partial_i \vect{m} =0$, which forces the magnetization to have 
no spatial gradients across the boundary. 
Remarkably, the internal DMI field $\vect{H}_D$ in Eq.~\eqref{Eq:Hd} is only determined by the antisymmetric part $D_{ijk} - D_{kji}$ of the DMI tensor, whereas
$\boldsymbol{\Gamma}_D$ depends on the full tensorial structure of the DMI. This means that the BCs are in general parametrized by more DMI parameters than the internal field.
Below, we explicitly show this for $C_{\infty v}$ systems, and demonstrate that the DMI can produce \emph{purely boundary-induced} spin textures that are independent of the internal DMI field. 

Systems with $C_{\infty v}$ symmetry are invariant under any proper or improper rotation about the high symmetry axis, which we in the present work assume to be along the $z$ direction.
The $C_{\infty v}$ symmetry  implies that the spin-stiffness matrix is determined by two independent tensor coefficients,
$J_{xx} = J_{yy}$ and  $J_{zz}$,
whereas four independent tensor coefficients govern the DMI tensor: 
$D_{xxz} = D_{yyz}$,  $D_{xzx} = D_{yzy}$, $D_{zxx} = D_{zyy}$,  and $D_{zzz}$.   
The remaining tensor coefficients vanish by symmetry.
It is common to parameterize these tensor coefficients by the antisymmetrized and symmetrized elements: 
\begin{subequations}
 \label{Eq:D}
\begin{align}
D^A_1\equiv (D_{zxx}-D_{xxz})/2, &\qquad D^S_2\equiv D_{xzx}, \\
D^S_1\equiv (D_{zxx}+D_{xxz})/2, &\qquad D^S_3\equiv D_{zzz}.
\end{align}
\end{subequations}
Here, superscript $S$ ($A$) labels the parts of $D_{ijk}$ that are symmetric (antisymmetric) with respect to the magnetic indices $i$ and $k$. 
Hence, the DMI free energy density $\mathcal{F}_D$, the internal DMI field $\vect{H}_D $, and the boundary-induced DMI field $\boldsymbol{\Gamma}_D$ are:
\begin{subequations}
\begin{align}
\mathcal{F}_D &= D_1^A \left[ m_z (\boldsymbol{\nabla}\cdot\vect{m} ) - \vect{m}\cdot\boldsymbol{\nabla} m_z \right]  \label{Eq:FdRashba} \\
	& \quad + (D_{23} - 2 D_1^S) m_z \partial_z m_z  + D_1^S \boldsymbol{\nabla}\cdot (m_z\vect{m}) , \nonumber  \\
\vect{H}_D &= 2 D_1^A \left[  \boldsymbol{\nabla} m_z - (\boldsymbol{\nabla} \cdot \vect{m})\hat{\vect{z}}  \right].  \label{Eq:Hd2}\\
\boldsymbol{\Gamma}_D &= (D_1^S + D_1^A) \vect{m}\times\left( \vect{n}\times\hat{\vect{z}} \right) +  D_2^S  n_z \vect{m} \\
& \quad + \left[  2 D_1^S (n_xm_x + n_ym_y) + D_{23} n_zm_z \right] \hat{\vect{z}}. \nonumber
\end{align}
\end{subequations}
Here, we have introduced $D_{23}=D^S_3 -D^S_2$, because $\mathcal{F}_D$ is governed by only three of the four independent tensor parameters due to $|\vect{m}|=1$.
Note that only the first term of $\mathcal{F}_D$, representing the standard N\'eel DMI~\cite{Thiaville2012,Rohart2013}, yields the internal DMI field $\vect{H}_D$. 
The latter is solely controlled by the antisymmetric tensor element $D_1^A$ and agrees with the effective DMI field considered in previous studies of $C_{\infty v}$ ferromagnets~\cite{Thiaville2012, Rohart2013, Wilson2013, Rybakov2013, Meynell2014, Leonov2015, Leonov2016c}.
The last two terms of $\mathcal{F}_D$ can be rewritten into boundary terms via the divergence theorem. The boundary terms do not contribute to the internal DMI field $\vect{H}_D $, but to the BC, which is characterized by all three independent parameters.  
Projecting out the component of $\boldsymbol{\Gamma}_D$ that is perpendicular to $\vect{m}$ yields the following BC for the magnetization: 
\begin{multline}
 2J_{ij}n_{i}\partial_{j}  \vect{m}  =  ( D_1^S + D_1^A ) \vect{m}\times\left( \hat{\vect{z}}\times\vect{n} \right) -   \\
   \left[ 2 D_1^S (n_xm_x + n_ym_y) +D_{23}  n_zm_z  \right] \vect{m}\times\left( \hat{\vect{z}}\times \vect{m} \right).
   \label{Eq:BondCondRashba}
   \end{multline}
Usually, the DMI in $C_{\infty v}$ ferromagnets is parameterized only by a single parameter ($D_1^A$ in our notation). While the antisymmetric parameter $D_1^A$ does give the correct internal DMI field [cf.\ Eq.~\eqref{Eq:Hd2}], this is not the case for the BC in Eq.~\eqref{Eq:BondCondRashba}. $D_1^A$ only produces parts of the term $\sim \vect{m}\times\left( \vect{n}\times\hat{\vect{z}} \right)$ in the surface field $\boldsymbol{\Gamma}_D $, whereas the component of $\boldsymbol{\Gamma}_D $ proportional to $\vect{m}\times\left( \hat{\vect{z}}\times \vect{m} \right)$ is solely an effect of the tensor coefficients $D_1^S$ and $D_{23}$. 
To conclude, we find that for $C_{\infty v}$ ferromagnets only one DMI parameter is necessary to capture the physics of the internal field, whereas three independent parameters are required to provide a correct micromagnetic description at the sample edges.

\begin{figure}[ht] 
\centering 
\includegraphics[scale=1.0]{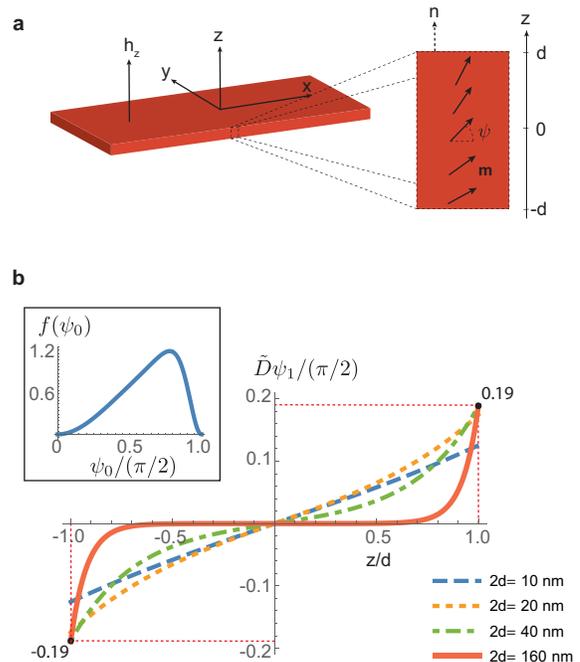}  
\caption{(Color online) \textbf{(a)} Illustration of the boundary-driven twist state. The black arrows 
in the close-up view of the sample indicate the canting of the magnetization across the ferromagnetic thin film. 
\textbf{(b)} The function $\tilde{D} \psi_1(z)$ for different thicknesses of the ferromagnetic film. At the boundaries (dotted vertical red lines) and for large thicknesses of the sample, $\tilde{D} \psi_1(z)$  approaches the $d\rightarrow \infty$ limit $0.19$ (dotted horizontal red lines). Inset: $\psi_0$-dependence of the texture-induced magnetoresistance for $2d=\SI{20}{\nano\meter}$. }
\label{Fig1} 
\end{figure}

We demonstrate below that the terms proportional to the symmetric DMI tensor coefficients in Eq.~\eqref{Eq:BondCondRashba} are not negligible and might lead to magnetic twist states that cannot be phenomenologically described without taking into account the full tensorial form of $D_{ijk}$. In particular, we show that there is a simple homogeneous ferromagnetic phase, where the usually exclusively considered antisymmetric part of the DMI vanishes in the BC, but the last term of Eq.~\eqref{Eq:BondCondRashba} induces a twist state. We estimate that the typical decay length of this twist state is about \SI{6}{\nano\meter}, which produces a magnetization gradient and an experimentally observable magnetoresistance effect that is larger than what is typically observed for magnetic domain walls. 
To study this new twist state, we consider an infinitely-large thin film of thickness $2d$ with the surface normal parallel to the high symmetry axis, see Fig.~\ref{Fig1}\textbf{a}. 
The magnet has an easy-plane anisotropy described by the free energy density $\mathcal{F}_a = K_u m_z^2$ where $K_u>0$. Additionally, it is subjected to an externally applied magnetic field along $z$, which produces the Zeeman energy $\mathcal{F}_h = h_z m_z$. 

The DMI is a relativistic effect and thus much smaller than the spin stiffness produced by the Coulomb interaction.
In the following, we assume that the DMI is small enough such that the magnetic system is mainly in a ferromagnetic phase with the magnetization tilted by an angle of $\psi$ out of the easy plane, see Fig.~\ref{Fig1}\textbf{a}. 
The system is translationally invariant in the $xy$-plane and rotational symmetric about $z$. Therefore, we can choose a coordinate system in which $\mathbf{m}$ lies in the $xz$-plane:  
\begin{equation}
\vect{m}(z) = (\cos \psi(z), 0, \sin \psi(z)), \label{Eq:Ansatz}
\end{equation}
and where the magnetization is fully determined by the tilt angle $\psi=\psi(z)$, which still might have a spatial variation along the high symmetry axis.
Because we consider an infinite film, the surfaces along the $x$ and $y$ axes can be disregarded and Eq.~\eqref{Eq:BondCondRashba} reduces to  
\begin{equation}
 2J_{zz}\partial_{z}  \vect{m}  =  
   D_{23} m_z  \left[ \vect{m}\times\left( \hat{\vect{z}}\times \vect{m} \right) \right].  
   \label{Eq:redboundary}
\end{equation}
Substituting Eq.~\eqref{Eq:Ansatz} into the free energy and minimizing the resulting energy functional with respect to $\psi$ results in the following equations for the interior and boundary of the sample:
\begin{subequations}
\label{Eq:BVP}
\begin{align}
\partial_{\tilde{z}}^{2} \psi &= \tilde{K}_u\sin (2\psi) - \tilde{h}_z\cos(\psi), \\
\partial_{\tilde{z}} \psi  |_{\tilde{z}=\pm 1}  &=  \tilde{D} \sin (2\psi).
\end{align}
\end{subequations}
Here, we have rescaled the $z$-axis by $\tilde{z}=z/d$, and introduced  
the dimensionless parameters $\tilde{K}_u\equiv  K_u d^2 / (2 J_{zz})$, $\tilde{h}_z\equiv  h_z d^2/ (2 J_{zz})$, and $\tilde{D}\equiv  D_{23} d / (4 J_{zz})$ controlling the strengths of the
anisotropy, external field, and boundary DMI, respectively.

Note that in our specific example $\vect{H}_D$ vanishes. Consequently, there are no magnetic textures produced by the equation for the inner part of the sample.
The solution of the full boundary value problem~\eqref{Eq:BVP} is, however, determined by the three dimensionless parameters $\tilde{K}_u$, $\tilde{h}_z$ and $\tilde{D}$ and does depend on the boundary DMI-field and the thickness of the film. In general, the thinner the film, the more the system is governed by the BC. 
In the following, we show that even in the case of larger thicknesses the influence of the BC is not negligible.

To treat the problem analytically, we assume in the following that $\tilde{D}$ is a small parameter and solve Eq.~\eqref{Eq:BVP} to first order in $\tilde{D}$.
To this end, we consider a solution of the form
\begin{equation}
\psi (\tilde{z}) = \psi_0 + \tilde{D} \psi_1 (\tilde{z}) , \label{Eq:AnsatzPsi}
\end{equation}
where $\psi_0$ is the constant solution for $ \tilde{D} = 0$, whereas $\psi_1$ represents the spatially varying DMI-induced correction to this solution. 
Substituting this ansatz into Eq.~\eqref{Eq:BVP}, we find that $\psi_0$ is given by 
\begin{equation}
 \sin (\psi_0)  = h_z/( 2 K_u) \quad \text{for} \quad |h_z/2K_u| < 1.
\end{equation}
$\psi_1$ is determined by the boundary value problem     
\begin{subequations}
\begin{align}
\partial_{\tilde{z}}^{2} \psi_1 &= \kappa_1 \psi_1  , \label{Eq:EqvEqPsi1} \\
\partial_{\tilde{z}} \psi_1 |_{\tilde{z}=\pm 1}  &=  \sin (2\psi_0) , \label{Eq:BVPPsi1}
\end{align}
\end{subequations}
where the dimensionless parameter $\kappa_1  = \kappa_1 (\psi_0) \equiv   2\tilde{K}_u\cos (2\psi_0) + \tilde{h}_z\sin(\psi_0)$  depends on the zeroth-order approximation of the tilting and on the thickness of the film, $\kappa_1 \sim d^2$.  
The analytical solution of the boundary value problem is
\begin{equation}
\psi_1 (\tilde{z}) = \frac{ {\sech} ( \sqrt{ \kappa_1} )  \sin (2\psi_0) }{ \sqrt{\kappa_1}} \sinh (\sqrt{\kappa_1} \tilde{z}).\label{Eq:SolPsi1}
\end{equation}
As expected, the modulation along the $z$ direction, $\psi_1(z)$, vanishes for $\psi_0 = 0$ and $|\psi_0|= \pi/2$ as $ m_z  \vect{m}\times\left( \hat{\vect{z}}\times \vect{m} \right)$ is zero in both cases.  

The functional form of $\tilde{D} \psi_1(\tilde{z})$ is shown in Fig.~\ref{Fig1} for different values of the film thickness $2d$.
Here, we have assumed the parameter values $J_{zz}= \SI{e-11}{\joule\per\meter}$, $K_{u} = \SI{5.1e5}{\joule\per\cubic\meter}$, $D_{23} = \SI{1.9e-3}{\joule\per\square\meter}$, and $h_z = \sqrt{2} K_u$, which are characteristic for ferromagnetic heterostructures \cite{Emori2013, Moreau-Luchaire2016}. In this case, the constant tilting angle induced by the the zeroth-order approximation for the inner part of the sample is $\psi_0 = \pi /4$. 

As the internal magnetic state is unaffected by the BCs in the limit of large film thickness,  $d\rightarrow \infty$, the correction induced by the boundaries of the sample, $\tilde{D} \psi_1(\tilde{z})$,  vanishes in the inner part of the sample as expected.
However, at the boundaries, $\tilde{D} \psi_1(\pm 1)$ does still lead to a correction and approaches the constant value $\pm \tilde{D} \sin (2\psi_0) / \sqrt{\kappa_1}\approx 0.19\times (\pi/2)$ for $d\rightarrow \infty$, see Fig.~\ref{Fig1}\textbf{b}.
Note that the parameter $\tilde{D} / \sqrt{\kappa_1}$  is independent of the thickness, because both $\tilde{D}$ and $\sqrt{\kappa_1}$ are linear in $d$.
The typical decay length of the twist state is $d/\sqrt{\kappa_1}\sim \SI{6}{\nano\meter}$, implying that the interface gives a significant correction to the internal spin structure in systems with thicknesses of \SI{12}{\nano\meter} or less.
For such thicknesses, the tilt angle approaches the solution $\psi \approx \psi_0 + \tilde{D} \sin(2\psi_0)\tilde{z}$, which represents a N\'eel helix (see Fig.~\ref{Fig1}\textbf{b}).    

The parameter $\lambda = \pi /|\tilde{D}\partial_{\tilde{z}} \psi_1|$ (evaluated at $\tilde{z}=\pm 1$) represents the strength of the magnetization gradient and corresponds to the width of a magnetic domain wall.
For the above parameter values, we find that $\lambda \sim \SI{7}{\nano\meter}$. 
This yields a magnetization gradient that is comparable to the typical spatial variations of magnetic domain walls (see Ref.~\citenum{Kent1999} for a review).

Since domain walls give observable corrections on the order of $1\%$ to the resistance of ferromagnetic systems \cite{Kent1999,Brataas1999a, Chiba2006}, we expect this also to be the case for the boundary-driven twist states discovered in this work. 
The texture-induced resistance has been extensively studied in the 
context of domain walls, where the spatially varying magnetization has been shown to produce a correction to the local resistivity proportional to $(\partial_i \vect{m})^2$ in the diffusive regime~\cite{Brataas1999a}.
In addition, there will be corrections from the anisotropic magnetoresistance (AMR) effect~\cite{Chiba2006}.
This contribution can be disregarded for currents applied perpendicular to the magnetic texture (\textit{i.e.}, along the $y$-axis in our case), because the AMR only depends on the relative angle between the applied current and the magnetization.  
Therefore, for currents along the $y$ direction we can phenomenologically model the local resistivity by
\begin{equation}
\rho (\tilde{z}) = \rho_0 +  \delta\rho_{\rm t} \left( \partial_{\tilde{z}} \psi (\tilde{z}) \right)^2 .\label{Eq:Resistivity}
\end{equation}
Here, $\rho_0$ is the resistance for currents perpendicular to the magnetization in the absence of any twist state, 
$ \delta\rho_{\rm t} $ parameterizes the resistivity caused by scattering at the magnetic texture,\cite{Brataas1999a} and $\left( \partial_{\tilde{z}} \psi (\tilde{z}) \right)^2 $ determines the magnitude of the magnetization gradient $(\partial_i \vect{m})^2$.

The total conductance of a system with lateral dimensions $L_\perp \times L_\| \times 2d$ is found by integrating ${\rm d}G =  [L_{\bot} d /(L_{||} \rho (\tilde{z}))] {\rm d\tilde{z}} $ over the thin film $\tilde{z}\in [-1,\ 1]$.
Here, $L_{||}$ ($L_{\bot}$) denote the length of the sample parallel (perpendicular) to the current. 
By substituting the perturbative solution~\eqref{Eq:AnsatzPsi} into the resistivity \eqref{Eq:Resistivity} and
assuming $|\delta \rho_{\rm t}/\rho_0| \ll 1$, one finds to linear order in $\delta \rho_{\rm t}/\rho_0$ the total conductance
\begin{equation}
G = \frac{1}{R_0} \biggl(1  -   \frac{\delta \rho_{\rm t} }{\rho_0} \frac{\tilde{D}^2 }{2} f(\psi_0)    \biggr)  , \label{Eq:G}
\end{equation}
where $R_0 = L_{||} \rho_0  / 2 d L_{\bot} $.
The second term represents the texture-induced correction with $f(\psi_0) \equiv \sin^2 (2\psi_0) {\sech}^2 (\sqrt{\kappa_1}) (1+  {\sinh} (2\sqrt{\kappa_1}) / 2 \sqrt{\kappa_1} )$, where $\kappa_1$ is a function of $\psi_0$ as before.

The magnetoresistance can be controlled by changing the tilt angle $\psi_0$, which experimentally can be manipulated by varying the strength of the applied magnetic field $h_z$. 
The $\psi_0$-dependence of the function $f$ is illustrated in the inset of Fig.~\ref{Fig1}\textbf{b}.   
The $f (\psi_0)$-modulation represents a clear signature of the texture-induced magnetoresistance effect, 
and can be used to distinguish this effect from other resistance phenomena in transport measurements. 
Based on previous works on domain wall resistance, the twist state is expected to produce a small but observable correction to the total resistance on the order of $1\%$.

To conclude, we have derived the boundary value problem for generic ferromagnets lacking spatial inversion symmetry. We have shown that the BCs require the full tensorial structure of the DMI tensor and not just the antisymmetric part.
We have specified the boundary value problem for ordinary systems with $C_{\infty v}$ symmetry.  Moreover, we have given an explicit example of a simple ferromagnetic thin film to demonstrate the importance of the correct BCs we have derived. Here, we have shown that the DMI leads to a purely boundary-driven magnetic twist state at the edges of the sample, which is completely independent of the internal DMI field. We have shown that such a twist state could be observed by conductance measurements upon varying the magnetic field. 
Overall, already our simple example highlights the importance of the correct treatment of the BCs and that it is notably important for the predictive power of micromagnetic simulations in confined geometries, specifically in those where boundary-induced effects might influence or even dominate the bulk properties.
 
We are grateful to M.~Sitte, A.~Bogdanov, A.~Leonov, M.~Garst and J.~Sinova for discussions. 
We further thank M.~Sitte for careful reading of the manuscript. 
We acknowledge the funding from the German Research Foundation (DFG) under the Project No.~EV~196/2-1.

\section*{Supplement}
The most general form of the DMI interaction is given by 
\begin{equation}
\mathcal{F}_d= D_{ijk}m_i\partial_j m_k,
\label{Eq:DMIintro}
\end{equation}
where the form of the DMI tensor is determined by the system's point group. In the literature, however, also different notations are used, which we briefly review here.

For example, it is common to symmetrize the DMI tensor with respect to the magnetic indices 
\begin{equation}
\mathcal{F}_d\equiv \frac{D_{(ik)j}}{2} \partial_j \left( m_i m_k \right) + \frac{D_{[ik]j}}{2} [ m_i\partial_j m_k - m_k\partial_j m_i ]  , \label{Eq:DMIintro2}
\end{equation}
where $D_{(ik)j}= (D_{ijk} + D_{kji})/2$ and  $D_{[ik]j}= (D_{ijk} - D_{kji})/2$ represent the symmetric and antisymmetric parts of the tensor, respectively. 
Other common representations of the DMI term are given by
\begin{align}
\mathcal{F}_d  &\equiv  \frac{D_{(ik)j} }{2}\partial_j(m_im_k)  +  \mathbf{D}_j\cdot (\partial_j \mathbf{m} \times \mathbf{m}),\\
                        &\equiv  \frac{D_{(ik)j} }{2}\partial_j(m_im_k)  + D_{ij} \mathcal{L}_{ij} (\mathbf{m}),
\end{align}
where $\mathbf{D}_j$, $D_{ij}=  (\mathbf{D}_j)_i$, and $ \mathcal{L}_{ij}=  (\partial_j \mathbf{m} \times \mathbf{m})_i$ are  
the Dzyaloshinskii-Moriya (DM) vectors, the spiralization tensor, and the chirality tensor, respectively. Furthermore, the relationship between the DM vectors and the DMI tensor is given by $(D_j)_{\mu} = -\epsilon_{\mu i k} D_{[ik] j}/2  $.

Note that upon a spatial integration, the symmetric part of the DMI interaction reduces to a surface term. Therefore, it is common to neglect $D_{(ik)j}$ and only keep the antisymmetric part -- described for example in terms of the DM vectors or the spiralization tensor -- which governs the DMI in the internal part of the sample. However, a correct description of the magnetization at the boundaries requires both the symmetric and antisymmetric part.

\end{document}